\documentstyle[twocolumn,aps,prl,epsf,floats]{revtex}
\begin{document}
\newcommand{\so}
	{\mathrel{\rlap{\raise1pt\hbox{$>$}}{\lower4pt\hbox{$\sim$}}}}
\newcommand{\bnu}
	{\mbox{\boldmath$\nu$}}
\draft

\twocolumn[\hsize\textwidth\columnwidth\hsize\csname@twocolumnfalse\endcsname

\title{Nematic-Wetted Colloids in the Isotropic Phase: Pairwise
Interaction,\\ Biaxiality and Defects}

\author{P. Galatola}
\address{LBHP, Universit\'e Paris 7---Denis Diderot, Case 7056, 2 place
Jussieu, F-75251 Paris Cedex 05, France}
\author{J.-B. Fournier}
\address{Laboratoire de Physico-Chimie Th\'eorique, E.\,S.\,P.\,C.\,I.,
10 rue Vauquelin, F-75231 Paris Cedex 05, France}
\date{\today}
\maketitle
\begin{abstract}
We calculate the interaction between two spherical colloidal particles
embedded in the isotropic phase of a nematogenic liquid. The surface of
the particles induces wetting nematic coronas that mediate an elastic
interaction. In the weak wetting regime, we obtain exact results for the
interaction energy and the texture, showing that defects and biaxiality
arise, although they are not topologically required. We evidence rich
behaviors, including the possibility of reversible colloidal aggregation
and dispersion. Complex anisotropic self-assembled phases might be
formed in dense suspensions.
\end{abstract}
\twocolumn]\narrowtext

Dispersions of small particles or liquid droplets in a host fluid,
namely colloidal suspensions, are a widespread and important state of
matter~\cite{Hunter}. They are long-lived metastable states, of
fundamental interest from the point of view of collective interactions
in complex matter. They also have considerable technological importance,
e.g., in paints, coatings, foods and drugs.  To prevent coagulation due
to attractive van der Waals forces, usually the particles are treated in
order to produce Coulombic or steric repulsive interactions. Recently, a
novel source of repulsion has been reported: the elastic distortion of a
liquid crystal host~\cite{Poulin97a}, i.e., a fluid phase with
long-range orientational order of the molecules. The repulsion arises from
the competition between the surface aligning property, which favors,
e.g., a radial orientation of the nematic molecules, and the bulk
elasticity, which favors a uniform nematic orientation~\cite{Poulin97b}.
This effect was shown to arise also in other anisotropic fluid hosts,
e.g., lyotropic solutions of anisotropic micelles~\cite{Poulin99}, and
in different liquid crystal phases, e.g.,
cholesterics~\cite{Zapotocky99}.  Lately, such systems were shown to
form liquid or solid composite materials with unusual
properties~\cite{Meeker00,Loudet00}.

It is well known that solid surfaces influence not only the orientation
of liquid crystals, but also, in general, the degree of order of their
constituent molecules. In particular, a surface can induce a wetting nematic
layer even above the transition temperature at which the nematic becomes
an ordinary isotropic liquid~\cite{Sheng76,Miyano79,Roij00}. Therefore,
an elastic colloidal stabilization could be achieved also above the
nematic--isotropic transition. In addition, the vicinity of a phase
transition may give a critical character to the stabilization mechanism
and yield rich phase-separation behaviors, as predicted
in~\cite{Lowen95} for a simpler system with a scalar order-parameter.  

To understand these complex collective effects, it is essential to
calculate the interaction between the particles precisely. An attempt,
based on a quasi-planar approximation and assuming uniaxiality of the
nematic tensorial order, has been proposed in Ref.~\cite{Borstnik99}.
Here, we give the first exact solution to the problem, including
biaxiality. Our calculation, based on a multipolar expansion, is valid
for weak surface ordering. We obtain numerically the texture between two
spherical particles imposing normal boundary conditions: unexpectedly,
we find a defect line that is not topologically required and a strong
biaxiality around it. We give an analytical expression for the
interaction energy at large distances, which we find always attractive.
At short distances, comparable to the nematic coherence length, we find
either attraction or repulsion, depending critically on the distance to
the nematic-isotropic transition. Collective behaviors could thus be
critically tuned in such systems, in analogy with, e.g., the switchable
tackiness or wettability of liquid crystal
polymers~\cite{deCrevoisier99}.

Nematic liquid crystals are anisotropic liquid phases, in which
elongated molecules display a long-range orientational order. Since
nematics are non-polar, this order is described by a symmetric traceless
tensorial order-parameter~$Q_{ij}$ ($i,j=1,2,3$). The eigenvectors
of~$Q_{ij}$ represent the axes of main molecular orientation and its
eigenvalues describe the amount of ordering and
biaxiality~\cite{deGennes}. Usually, nematics are uniaxial phases,
however biaxiality naturally arises in inhomogeneous situations, e.g.,
in the vicinity of defects~\cite{Kralj99}. For a weak induced nematic
order in the isotropic phase, the quadratic Landau-de Gennes expansion
of the bulk free-energy density~\cite{deGennes} has the form
\begin{equation}\label{eq:Landau}
f = \frac{1}{2} a\, Q_{ij} Q_{ij} + \frac{1}{2} L\, Q_{ij,k} Q_{ij,k},
\end{equation}
were comma indicates derivation and summation over repeated indices is
implied. Here $a>0$ (resp.\ $L>0$) quantifies the cost of creating
(resp.\ distorting) the nematic phase. For simplicity, we take a
one-constant approximation for the gradient terms. We consider spherical
colloidal particles dispersed in the isotropic phase of the nematic.
They favor a uniaxial nematic order $Q_{ij}^{(0)} = S_0 (\nu_i \nu_j -
\frac{1}{3} \delta_{ij})$ on their surface, of outward normal~$\bnu$.
Here, $0\le S_0\le 1$ is the preferred surface scalar order parameter
and $\delta_{ij}$ the Kronecker delta. At quadratic order, the
corresponding surface free-energy density is
\begin{equation}
f_s = \frac{1}{2} W \left( Q_{ij} - Q_{ij}^{(0)}\right)
\left( Q_{ij} - Q_{ij}^{(0)}\right),
\end{equation}
where $W$ measures the anchoring strength.

Scaling lengths to the nematic correlation length $\xi=(L/a)^{1/2}$ and
energies to $F_0=a\xi^3 S_0^2$, the total free-energy can be written as
\begin{equation}
F = \sum_{i=0}^4\! c_i\!\left\{
\int\!\! \left[ q_i^2 + \left(\nabla q_i\right)^2 \right]\!
d^3 r
+ w \!\!\int\!\!\! \left(q_i-q_i^{(0)}\right)^2\! d^2 r
\right\},
\end{equation}
where $q_0 = (Q_{xx} - Q_{yy})/S_0$, $q_1 = Q_{yz}/S_0$, $q_2 =
Q_{xz}/S_0$, $q_3 = Q_{xy}/S_0$, $q_4 = Q_{zz}/S_0$, $c_0 = 1/4$,
$c_1 = c_2 = c_3 = 1$, $c_4 = 3/4$, and $w = W/(a\xi)$ is a
normalized anchoring strength. The corresponding equilibrium equations are
\begin{equation}
\label{eq:bulk}
\nabla^2 q_i = q_i,
\end{equation}
in the bulk, and
\begin{equation}
\label{eq:surface}
\bnu\cdot\nabla q_i = w\left(q_i - q_i^{(0)}\right),
\end{equation}
on the surface of each colloidal particle. In spherical coordinates
$(r_p,\theta_p,\phi_p)$ centered on colloidal particle~$p$, the
general solution of Eq.~(\ref{eq:bulk}) that is regular everywhere but
in $r_p=0$ can be written as a multipolar expansion:
\begin{equation}
\label{eq:multi}
q_i = \sum_{\ell=0}^\infty \sum_{m=-\ell}^\ell q_{ip}^{\ell m} u_\ell(r_p)
Y_{\ell m}(\theta_p,\phi_p),
\end{equation}
where $Y_{\ell m}(\theta_p,\phi_p)$ are spherical harmonics and
$u_\ell(r_p) = \sqrt{2/(\pi r_p)} K_{\ell+1/2}(r_p)$, in terms of
half-integer modified Bessel functions. The solution to Eqs.\
(\ref{eq:bulk}) and~(\ref{eq:surface}) can be written as a superposition
of a multipolar expansion~(\ref{eq:multi}) for each particle, the
coefficients of which can be adjusted by imposing the boundary
conditions~(\ref{eq:surface}) for each spherical harmonic. 

In the following, we consider the case of two identical particles, with
reduced radius~$R$ and centers in $z=\pm d/2$. Then, by symmetry,
\begin{mathletters}
\begin{eqnarray}
q_0 &=& \alpha(r,\theta) \cos 2\phi,\\
q_1 &=& \beta(r,\theta) \sin\phi,\\
q_2 &=& \beta(r,\theta) \cos\phi,\\
q_3 &=& \case{1}{2} \alpha(r,\theta) \sin 2\phi,\\
q_4 &=& \gamma(r,\theta),
\end{eqnarray}
where $(r,\theta,\phi)$ are spherical coordinates relative to the
origin, and
\end{mathletters}
\begin{mathletters}
\label{eq:expansion}
\begin{eqnarray}
\alpha(r,\theta) &=& \sum_{\ell=2}^\infty \sum_{p=1}^2 \alpha_\ell
u_\ell(r_p) P_\ell^2(\cos\theta_p), \\
\beta(r,\theta) &=& \sum_{\ell=1}^\infty \sum_{p=1}^2 (-1)^p \beta_\ell
u_\ell(r_p) P_\ell^1(\cos\theta_p),\\
\gamma(r,\theta) &=& \sum_{\ell=0}^\infty \sum_{p=1}^2 \gamma_\ell
u_\ell(r_p) P_\ell^0(\cos\theta_p),
\end{eqnarray}
in which the $P_\ell^m$ are the modified Legendre functions appearing in
the spherical harmonics~$Y_{\ell m}$.
\end{mathletters}

\begin{figure}
\centerline{\epsfxsize=7cm\epsfbox{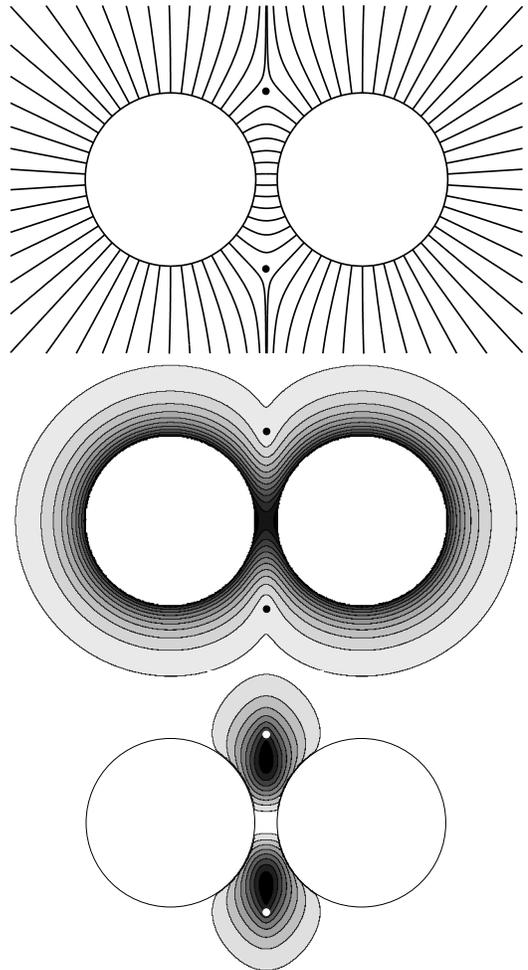}}
  \caption{Order-parameter profile between two spheres of reduced radius
  $R=2$ with normalized anchoring strength $w=4$. The profile has
  cylindrical symmetry about the axis joining the centers of the two
  spheres. Top: field lines of the nematic director~$\bf n$; Middle:
  contour lines of the uniaxial order parameter~$S$; Bottom: contour
  lines of the biaxial order parameter~$B$. The dots indicate the center
  of the core (the biaxial region) of the $-{1\over 2}$ defect Saturn ring.}
\label{fig:textures}
\end{figure}

Solving for $\alpha_\ell$, $\beta_\ell$, $\gamma_\ell$ at leading order
in the reduced separation~$d$, we obtain the asymptotic interaction
free-energy (subtracted of the self-energy of each particle):
\begin{equation}
\label{eq:asym}
F \sim \frac{-8\pi w^2 R^8 \exp(-d+2R)}{3d\left[9+3\left(3+w\right)R +
\left(4+3w\right) R^2 +\left(1+w\right) R^3\right]^2}.
\end{equation}
Hence, the interaction is always asymptotically attractive, in
disagreement with the estimation of Ref.~\cite{Borstnik99}.

At arbitrary distance~$d$, we solve numerically the linear system giving
$\alpha_\ell$, $\beta_\ell$, $\gamma_\ell$ by truncating the
expansions~(\ref{eq:expansion}) at some order $\ell_{\mathrm max}$ and
checking for convergence. From the three eigenvalues $\lambda_1 >
\lambda_2 > \lambda_3$ (with $\lambda_1+\lambda_2+\lambda_3=0$) we
obtain the scalar order-parameter $S=\frac{3}{2}\lambda_1$ and the
biaxiality $B=\frac{1}{2}(\lambda_2-\lambda_3)$. We call nematic
director $\mathbf n$ the direction corresponding to the largest
eigenvalue~$\lambda_1$.

Typical profiles of $\mathbf n$, $S$ and~$B$ are shown in
Fig.~\ref{fig:textures}. The tensor $\mathsf Q$ is everywhere
continuous, however, a defect (discontinuity) of~$\mathbf n$ appears, in
the form of a ``Saturn ring'' of strength $-{1\over2}$. This defect is
not topologically required: one can expand it to infinity, where it
would disappear since $\mathsf Q$ vanishes. Thus, it arises
spontaneously to minimize the energy, similarly to focal conic defects
in smectic-$A$ ``bat\^onnets''~\cite{JBF91}. Note that our calculation
does include the defect's core energy, since it resolves the
gradients of $\mathsf Q$ also within the core. Our only
approximation is the neglect of higher-order gradient terms in
Eq.~(\ref{eq:Landau}): this is however justified as long as $\xi$ is
much larger than the range of molecular interactions. Biaxiality
develops within the Saturn ring. Again, this is surprising, since both
the bulk and the surface favor uniaxiality. Thus, any modeling of
similar systems assuming uniaxiality might miss important features. The
behavior of the eigenvalues of~$\mathsf Q$ along a diameter of the
defect is shown in Fig.~\ref{fig:lambda}.  On the ring's center,
$\lambda_1=\lambda_2$: the nematic is locally uniaxial in the direction
of the ring, however with a discotic-like ordering, since the ring axis
corresponds to the smallest eigenvalue. 

\begin{figure}
\centerline{\epsfxsize=7cm\epsfbox{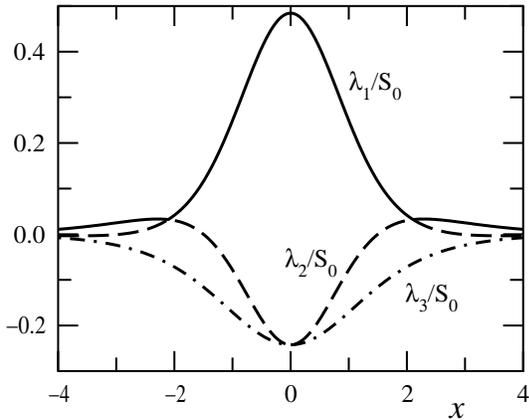}}
  \caption{Eigenvalues of the nematic order parameter $\sf Q$ along a
  diameter of the $-{1\over2}$ defect Saturn ring whose section is shown in
  Fig.~\protect\ref{fig:textures}. On the ring's center, $\lambda_1$
  and~$\lambda_2$ coincide.}
\label{fig:lambda}
\end{figure}

The behavior of the interaction energy as a function of the
separation~$d$ exhibits many detailed features. At large $d$'s, it is
always attractive, in agreement with the asymptotic
expression~(\ref{eq:asym}). Then, we distinguish six
scenarios, depending on the value of $R$ and~$w$ (Fig.~\ref{fig:phase}).
In case $0$, the interaction is monotonically attractive. In case $1a$
(resp.\ $1b$), a minimum develops, the interaction energy at contact
being negative (resp.\ positive), as shown in Fig.~\ref{fig:forme1}. In
case $2a, 2b$ and $2c$, the interaction energy displays first a minimum
then a maximum, as $d$ decreases. In case $2a$, both the maximum and the
contact energy are negative; in case $2b$, the maximum is positive while
the contact energy is negative; finally, in case $2c$ both the maximum
and the contact energy are positive (Fig.~\ref{fig:forme2}).

\begin{figure}
\centerline{\epsfxsize=7cm\epsfbox{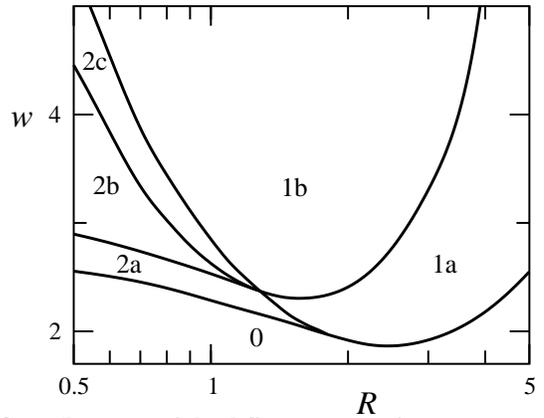}}
  \caption{Diagram of the different types of interaction energy
  profiles.}
\label{fig:phase}
\end{figure}

\begin{figure}
\centerline{\epsfxsize=7cm\epsfbox{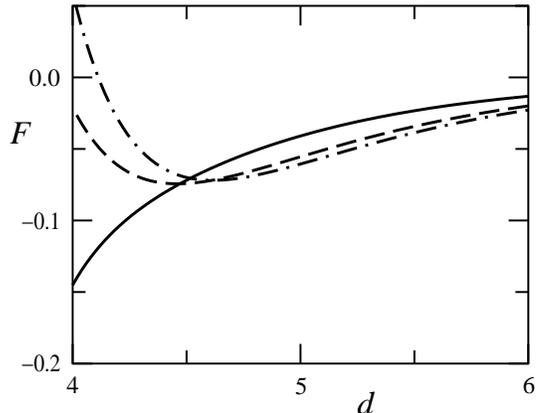}}
  \caption{Dimensionless interaction energy~$F$ as a function of $d$ for
  $R=2$ and $w=1.5$ (continuous line: case $0$), $w=2.3$ (dashed line:
  case $1a$), $w=2.7$ (dash-dotted line: case $1b$).}
\label{fig:forme1}
\end{figure}

\begin{figure}
\centerline{\epsfxsize=8cm\epsfbox{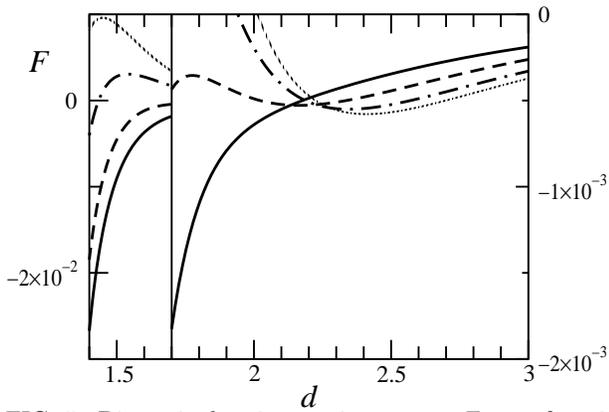}}
  \caption{Dimensionless interaction energy~$F$ as a function of $d$ for
  $R=0.7$ and $w=2$ (continuous line: case $0$), $w=2.6$ (dashed line:
  case $2a$), $w=3.2$ (dash-dotted line: case $2b$), $w=3.6$ (dotted
  line: case $2c$).  Different scales for $F$ are used on the two sides
  of the vertical line.}
\label{fig:forme2}
\end{figure}

To understand the collective behavior of such colloids, one needs to
compare the amplitudes of the barriers and of the wells of the
interaction energy with $k_{\mathrm B}T$. To simplify, let us assume
that the colloids are treated, e.g., with surfactants or polymer
coatings, such as to prevent aggregation due to van der Waals
attraction.  Qualitatively, we distinguish three different cases.
Whenever the interaction energy displays a minimum deeper than
$k_{\mathrm B}T$, attraction occurs, yielding aggregation either at
contact or at a finite distance.  Otherwise, two cases are possible. If
a maximum is present with a height larger than $k_{\mathrm B}T$,
short-range repulsion occurs: the nematic alone could stabilize the
colloidal dispersion. If, on the other hand, the energy is everywhere
smaller than $k_{\mathrm B}T$, the interaction is indifferent.
Figure~\ref{fig:diagram} displays the phase behavior of colloidal
particles of fixed radius when changing the distance to the nematic
transition. Here we have assumed $L\simeq k_{\mathrm B}T/m$, $S_0^2
\simeq 0.5$, and $W\simeq k_{\mathrm B}T/m^2$, where $m$ is a molecular
or a micellar length.  Note that $W$ is a microscopic anchoring,
stronger than the corresponding coarse-grained anchoring that would be
measured at a macroscopic scale~\cite{JBF99}.

In conclusion, we have investigated the interaction between colloidal
particles dispersed in the isotropic phase of a nematogenic liquid. Our
calculations show that small particles, of size ranging from a few to
several hundreds nematic correlation lengths, could be reversibly
dispersed or aggregated by tuning the vicinity to the nematic
transition, e.g., by changing the temperature.
Interesting effects might occur when more than two particles are brought
together. Standard aggregation favors three-dimensional close packing.
Here, however, the onset of extra defects might produce anisotropic
self-assembled structures~\cite{safran2000}, although both the embedding
fluid and the particles are isotropic. Indeed, while it is obvious that
three particles in a line will create two Saturn rings, it is not easy
to picture the defects, and the associated energy, in the case of three
particles forming a triangle, or four forming a tetrahedra. Our method
allows to calculate the textures and free-energies of arbitrary
arrangements of colloidal particles. Such detailed studies are however
beyond the scope of this paper, since they require heavy
numerical simulations: further analyses are required to investigate
whether gel-like structures (when linear aggregates are favored) or
smectic-like structures (when two-dimensional close-packing is favored)
could arise~\cite{progress}.

P. G. acknowledges financial support from ESPCI through a Joliot
visiting chair.

\begin{figure}
\centerline{\epsfxsize=7cm\epsfbox{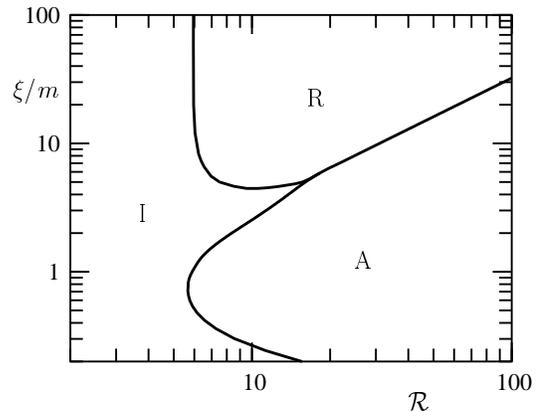}}
  \caption{Different regimes of the interaction between two colloidal
particles, as gauged by~$k_{\mathrm B}T$: $\mathrm I$ indifferent,
$\mathrm R$ repulsive, $\mathrm A$ attractive. $\mathcal R$ is the
particle radius in molecular or micellar units~$m$, and $\xi$ is the
nematic coherence length.}
\label{fig:diagram}
\end{figure}


\end{document}